\documentclass[aps,prb,showpacs,groupedaddress,twocolumn]{revtex4}
\usepackage{amsmath,amssymb}
\usepackage{graphicx} 
\usepackage{dcolumn} 
\usepackage{bm} 

\begin{document}
\draft
\title{New universal conductance fluctuation of mesoscopic systems in the
crossover regime from metal to insulator}
\author{Zhenhua Qiao, Yanxia Xing, and Jian Wang$^{*}$}
\address{Department of Physics and the Center of Theoretical and
Computational Physics, The University of Hong Kong, Hong Kong,
China}


\begin{abstract}
We report a theoretical investigation on conductance fluctuation of
mesoscopic systems. Extensive numerical simulations on quasi-one
dimensional, two dimensional, and quantum dot systems with different
symmetries (COE, CUE, and CSE) indicate that the conductance
fluctuation can reach a new universal value in the crossover regime
for systems with CUE and CSE symmetries. The conductance fluctuation
and higher order moments vs average conductance were found to be
universal functions from diffusive to localized regimes that depend
only on the dimensionality and symmetry. The numerical solution of
DMPK equation agrees with our result in quasi-one dimension. Our
numerical results in two dimensions suggest that this new universal
conductance fluctuation is related to the metal-insulator
transition.
\end{abstract}
\maketitle

One of the most important features in mesoscopic system is that the
conductance in the diffusive regime exhibits universal features with
a universal conductance fluctuation (UCF) that depends only on the
dimensionality and the symmetry of the system.\cite{lee85} There
exist three ensembles or symmetries according to the random matrix
theory (RMT)\cite{beenakker}: (1). Circular orthogonal ensemble
(COE) (characterized by symmetry index $\beta=1$) when the
time-reversal and spin-rotation symmetries are present. (2).
Circular unitary ensemble (CUE) ($\beta=2$) if time-reversal
symmetry is broken. (3). Circular symplectic ensemble (CSE)
($\beta=4$) if the spin-rotation symmetry is broken while
time-reversal symmetry is maintained. In the diffusive regime, the
UCF is given by $rms (G) = c_d /\sqrt{\beta} e^2/h$ where $c_d
=0.73, 0.86, 0.70$ for quasi-one dimension (1D), two dimensions (2D)
and quantum dot (QD) systems and
$\beta=1,2,4$.\cite{lee85,beenakker} Although the RMT can apply to
both diffusive and localized regimes, so far the universal
conductance fluctuation has been addressed and established only in
the diffusive regime. When the system is away from the diffusive
regime, some universal behaviors have been observed. For instance,
the conductance distribution of quasi-1D systems ($\beta=1$) with
surface roughness was found to be universal in the crossover regime,
independent of details of the system.\cite{saenz1} For quasi-1D
systems with $\beta=1,2$ the conductance distribution obtained from
tight-binding model agrees with the numerical solution of DMPK
equation.\cite{saenz2} In the localized regime, the conductance
distribution of the quasi-1D system obeys log normal
distribution.\cite{wolfle} In high dimensions, conductance
distribution at the mobility edge of metal-insulator transition was
also shown numerically to be universal for 2D systems with
$\beta=2,4$ and a 3D system with $\beta=1$.\cite{soukoulis} In the
localized regime, the conductance distribution of 3D systems is
qualitatively different from that of quasi-1D systems.\cite{wolfle1}
It would be interesting to further explore the universal behaviors
of these systems and ask following questions: Is there any universal
behaviors away from mobility edge? Is it possible to have a UCF
beyond the diffusive regime? If there is, what is the nature of the
new UCF? It is the purpose of this work to investigate these issues.

To do this, we have carried out extensive numerical calculations for
conductance fluctuations $rms(G)$ in quasi-1D, 2D and QD systems for
different symmetries: $\beta=1,2,4$. Our results can be summarized
as follows.
(1). From diffusive to localized regimes, the conductance
fluctuation and higher order moments were found to be universal
functions of the average conductance for quasi-1D, 2D, and QD
systems and for $\beta=1,2,4$. (2). We found that there exists a
second UCF near the localized regime for $\beta=2,4$ but not for
$\beta=1$. Our results show that the new UCF depends weakly on the
symmetries of the system and assumes the following value: $rms (G) =
{\tilde c}_d e^2/h$. Here for $\beta=2$ we found ${\tilde c}_d =
0.56 \pm 0.01, 0.68 \pm 0.01, 0.58 \pm 0.01$ for quasi-1D, 2D and QD
systems, respectively while for $\beta=4$ we have ${\tilde c}_d =
0.55 \pm 0.01, 0.66 \pm 0.02, 0.56 \pm 0.01$. The conductance
distribution in this new regime was found to be one-sided log-normal
in agreement with previous results.\cite{wolfle} (3). In the
localized regime with $\left\langle G \right\rangle<0.3$, the
conductance distribution does not seem to depend on dimensionality
and symmetry. (4). For quasi-1D systems, the numerical solution of
DMPK equation\cite{dmpk} agrees with our results. (5). For quasi-1D
systems, the new UCF occurs when the localization length $\xi$ is
approximately equal to the system size $L$, i.e., $\xi \sim L$ for
$\beta=2,4$. For 2D systems, we found that the new UCF occurs in the
vicinity of the critical region of metal-insulator transition.

\begin{figure}
\includegraphics[width=9cm,totalheight=6.5cm,angle=0]{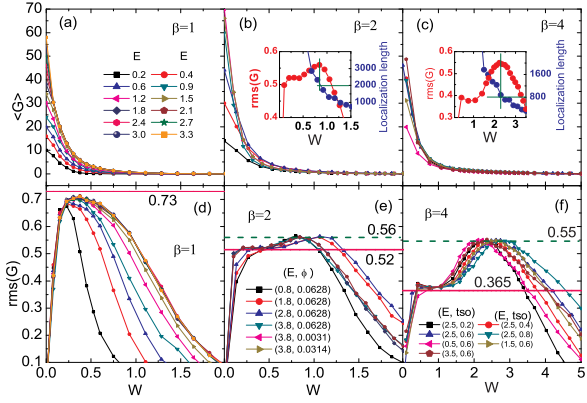}
\caption{(Color online) Average conductance (a,b,c) and its
fluctuation (d,e,f) vs disorder strength $W$ for different symmetry
index $\beta$ in quasi-1D systems. Insets: localization length vs
$W$ for $\beta=2,4$.} \label{fig1}
\end{figure}

In the numerical calculations, we used the same tight-binding
Hamiltonian as that of Ref.\onlinecite{qiao}. Static Anderson-type
disorder is added to the on-site energy with a uniform distribution
in the interval $[-W/2,W/2]$ where $W$ characterizes the strength of
the disorder. The conductance $G$ is calculated from the
Landauer-Buttiker formula $G=({2e^2}/{h}) T$ with $T$ the
transmission coefficient. The conductance fluctuation is defined as
$\text{rms}(G)\equiv \sqrt{\left\langle G^{2}\right\rangle
-\left\langle G\right\rangle ^{2}}$, where $\left\langle
{\cdots}\right\rangle $ denotes averaging over an ensemble of
samples with different disorder configurations of the same strength
$W$. In the following the average conductance and its fluctuation
are measured in unit of $e^2/h$; the magnetic field is measured
using magnetic flux $\phi=\mu_B B/t$ where $\mu_B$ is Bohr magneton
and $t$ is the hopping energy that is used as the unit of energy.

We first examine average conductances $\left\langle G \right\rangle$
and their fluctuations $rms(G)$ vs disorder strength $W$ in quasi-1D
systems with different symmetry index $\beta$ (see Fig.1). In our
numerical simulation, the size of quasi-1D systems are chosen to be
$40 \times 2000$ for $\beta=1,2$ (Fig.1a,b) and $40 \times 800$ for
$\beta=4$ (Fig.1c). Each point in Fig.1 is obtained by averaging
9000 configurations for $\beta=1$ and 15000 configurations for
$\beta=2,4$. In Fig.1, data with different parameters are shown. For
instance with $\beta=2$, $\left\langle G \right\rangle$ and $rms(G)$
vs $W$ are plotted for different Fermi energies with fixed
$\phi=0.0628$ and different magnetic flux with fixed Fermi energy
$E=3.8$. From Fig.1, we see that in the diffusive regime where
$\left\langle G \right\rangle > 1$ there is a plateau region for
$rms(G)$ with the plateau value approximately equal to the known UCF
values $rms (G) = 0.73 /\sqrt{\beta}=0.73, 0.52, 0.365$ (marked by
solid lines). This suggests that one way to identify UCF is to
locate the plateau region in the plot of $rms(G)$ vs disorder
strength and the plateau value should correspond to UCF. This method
has been used to identify universal spin-Hall conductance
fluctuation\cite{ren} that was later confirmed by RMT.\cite{confirm}
Importantly, there exists a second plateau region for $\beta=2,4$
but not for $\beta=1$. The new plateau value approximately equals to
$rms (G)=0.56 \pm 0.01$ for $\beta=2$ and $rms (G)=0.55 \pm 0.01$
for $\beta=4$. In this regime, we found that $\left\langle G
\right\rangle \le 1$ which clearly corresponds to the crossover
regime. To confirm that the first and second plateaus are indeed in
the diffusive and crossover regimes, respectively, we have
calculated the localization length $\xi$ of the quasi-1D system
(insets of Fig.1). It is clear that near the first plateau where $W
\sim 0.4$ for $\beta=2$ and $W \sim 1$ for $\beta=4$ we have $\xi
\gg L$ with $L$ the length of quasi-1D system while near the second
plateau we have $\xi \sim L$ (see insets of Fig.1).

\begin{figure}
\includegraphics[width=9cm,totalheight=6.5cm,angle=0]{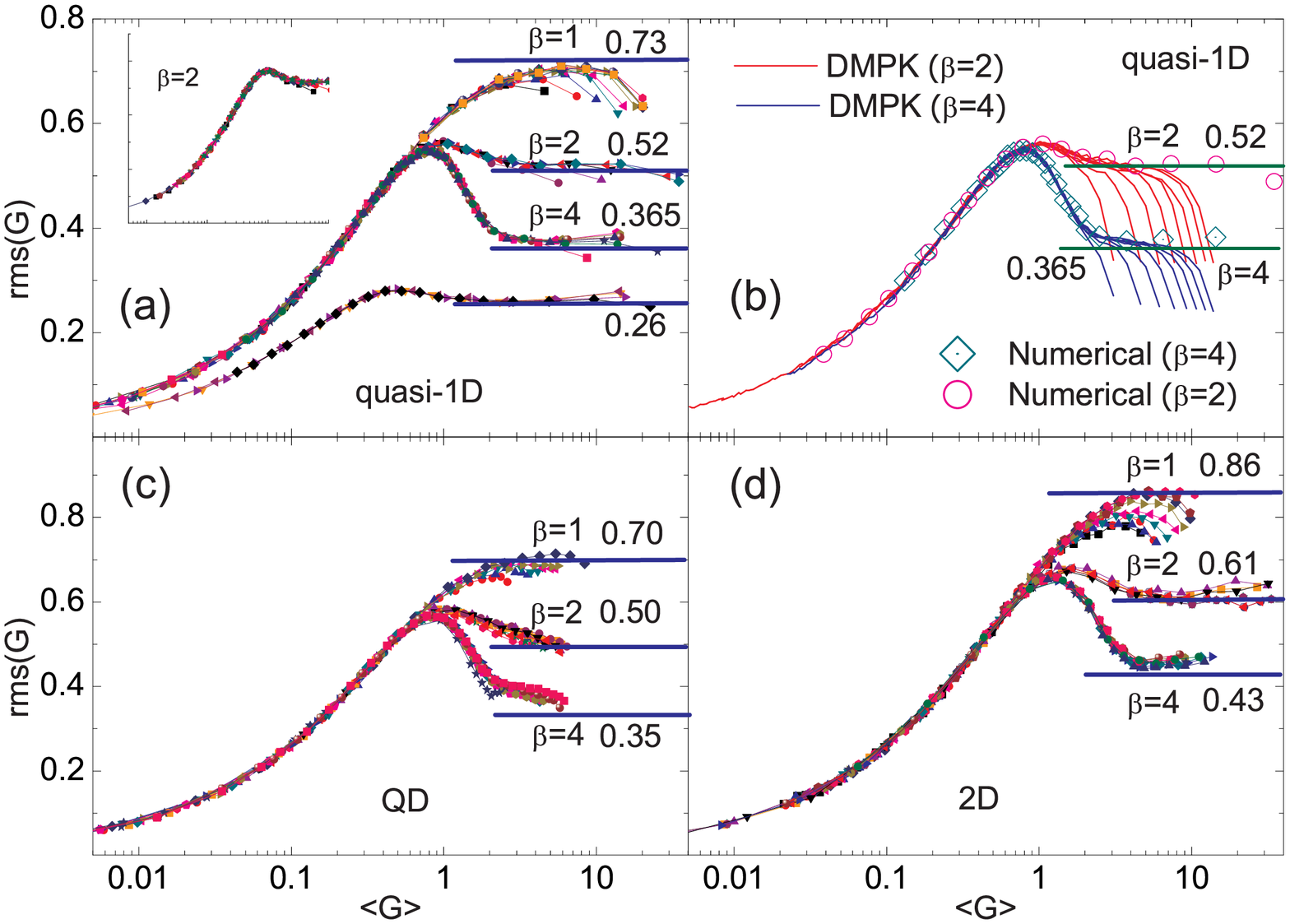}
\caption{(Color online) Conductance fluctuation vs average
conductance for quasi-1D systems (a,b), QD systems (c), and 2D
systems (d).} \label{fig2}
\end{figure}

According to the UCF in the diffusive regime, it is tempting to
conclude that this second plateau should correspond to a new UCF.
However, in making such a claim one has to answer following
questions: (1). whether the plateau behavior is universal? (2). if
it is, whether it can be observed in a wide range of parameters?
(3). how is our result compared with the theoretical predictions
whenever available? (4). whether such a universal behavior exists in
high dimensions? In the following, we provide evidences that the new
plateau indeed corresponds to a new UCF.

To answer the first question, we plot $rms(G)$ vs $\left\langle G
\right\rangle$ in Fig.2a by eliminating $W$. The fact that all
curves shown in Fig.1 with different parameters (Fermi energy $E$,
magnetic flux $\phi$, and SOI strength $t_{so}$) collapse into a
single curve for each $\beta$ strongly indicates that $rms(G)$ vs
$\left\langle G \right\rangle$ is universal. To further demonstrate
this universal behavior, we have calculated the conductance
fluctuation for a quasi-1D system with both magnetic flux and Rashba
SOI. Although the Hamiltonian of this system is still unitary, both
time reversal and spin rotation symmetries are broken. According to
the diagrammatic perturbation theory\cite{feng}, the UCF is reduced
by a factor of 2 when SOI is turned on. From RMT point of view, both
systems ($B \ne 0$, $t_{so}=0$) and ($B \ne 0$, $t_{so} \ne 0$) are
unitary ensembles and obey the same statistics. The fact that energy
spectrum for ($B \ne 0$, $t_{so}=0$) is doubly degenerate accounts
for the factor of 2 reduction for the system ($B \ne 0$, $t_{so} \ne
0$). In Fig.2a, we have plotted $rms(G)$ vs $\left\langle G
\right\rangle$ for the system with ($B \ne 0$, $t_{so} \ne 0$). Once
again, we see that all data from different parameters collapse into
a single curve. If we multiply this curve by a factor of 2, it
collapses with the curve of $\beta=2$ (see inset of Fig.2a).

\begin{figure}
\includegraphics[width=9cm,totalheight=6.5cm,angle=0]{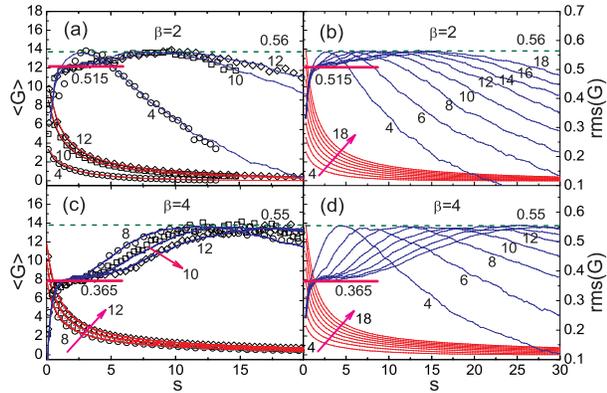}
\caption{(Color online) Average conductance and its fluctuation vs s
for quasi-1D systems [(a) and (c)] and compared with results of DMPK
equation for different $N=4,6,...,18$ [(b) and (d)]. The arrow
points the direction of increasing $N$.} \label{fig3}
\end{figure}

From Fig.2a we see that in the diffusive regime with $\left\langle G
\right\rangle > 1$ the long plateau for each $\beta$ corresponds to
the known UCF (marked by solid lines). For $\beta=1$ there is only
one plateau. For $\beta=2,4$, however, a second plateau region is in
the neighborhood of $\left\langle G \right\rangle \sim 1$. For
$\beta=2$, $rms(G)$ is approximately a constant hence {\it
universal} in the region $\left\langle G \right\rangle=(0.6,1.4)$
while for $\beta=4$ this region narrower with $\left\langle G
\right\rangle=(0.7,0.9)$. Looking at Fig.2a, it seems that the
second plateau region is narrower than the first one. But if we look
at $rms(G)$ vs $W$ where $W$ can be controlled experimentally, the
crossover region is enlarged since in the crossover regime
$\left\langle G \right\rangle$ is not very sensitive to $W$ while in
the diffusive regime it is the opposite. Indeed, in Fig.1f, we do
see that the ranges of the first and second plateau regions are
comparable. If we fix $W$ and plot $\left\langle G \right\rangle$
and $rms(G)$ vs the length of the system, the crossover regime is
enlarged further. These results are shown in Fig.3a,c where the
symbols represent our numerical result and solid lines correspond to
exact solution of DMPK equation (to be discussed below). We see that
the window of the second plateau is much larger than the first one.

Since the statistics of transmission eigen-channels of quasi-1D
systems can be described by DMPK equation, we have numerically
solved it\cite{shi} for $\beta=2,4$ and compared our numerical
results of tight-binding model with that of DMPK. Fig.3b and Fig.3d
show the numerical solution of DMPK equation for $\beta=2,4$ with
different $N=4,6,...,18$ where $N$ is the number of transmission
channels. Here $s=L/{\bar l}$ where $L$ is the length of quasi-1D
systems and ${\bar l}$ is the average mean free
path.\cite{beenakker,new} Fig.3 clearly shows that in the diffusive
regime where $1\ll s \ll N$ the first plateau corresponds to the
usual UCF and there exists a much wider second plateau in the
crossover regime where $s$ and $N$ are comparable with plateau value
equal to our newly identified UCF. Fig.3a and Fig.3c show the
comparison between the results of DMPK and that of quasi-1D tight
binding model. The $rms(G)$ vs $\left\langle G \right\rangle$ of
DMPK equation is plotted in Fig.2b where selected data from Fig.2a
is also plotted for comparison. The agreement between numerical and
theoretical results is clearly seen.

\begin{figure}
\includegraphics[width=9cm,totalheight=6.5cm,angle=0]{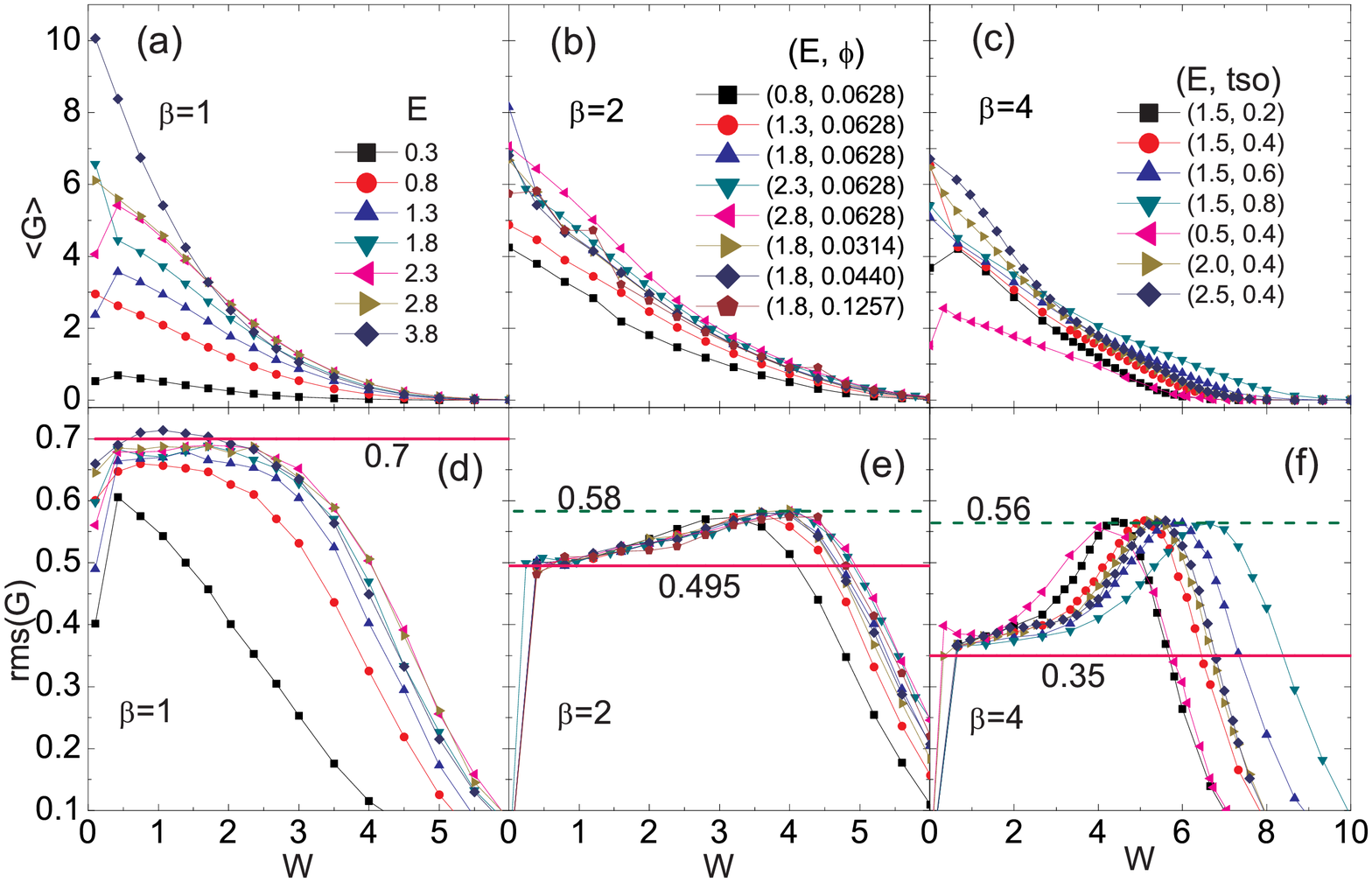}
\caption{(Color online) Average conductance and its fluctuation vs
$W$ for QD systems with $\beta=1,2,4$.} \label{fig4}
\end{figure}

Now we examine the conductance fluctuation for QD systems. In the
numerical calculation, the size of QD is $L \times L$ with $L=100$
and the width of the lead $L_0=10$ for $\beta=2,4$ while for
$\beta=1$ we used $L=150$. Fig.4 depicts $\left\langle G
\right\rangle$ and $rms(G)$ vs $W$ for $\beta=1,2,4$. Each point in
Fig.4 was obtained by averaging 9000 configurations for $\beta=1$
and 20000 configurations for $\beta=2,4$. Similar to quasi-1D
systems, we see only one plateau region in the diffusive regime for
$\beta=1$ with plateau value close to the known UCF value
$rms(G)=0.70$. In addition to the first plateau in the diffusive
regime, there exists a second plateau for $\beta=2,4$ which we
identify to be the regime for new UCF. The new UCF is again in the
crossover regime where $\left\langle G \right\rangle \sim 1$ with
the value ${\rm UCF}(\beta=2) = 0.58 \pm 0.01$ and ${\rm
UCF}(\beta=4) = 0.56 \pm 0.01$. In Fig.2c we plot $rms(G)$ vs
$\left\langle G \right\rangle$. It shows that all curves for each
$\beta$ collapse into a single curve showing universal behaviors.
Fig.4e and Fig.4f show that the new universal regime can be quite
large. Finally we have also calculated the conductance fluctuation
for 2D systems and similar behaviors were found (see
Fig.2d).\cite{foot4} In particular, the values of new UCF are found
to be ${\rm UCF}(\beta=2) = 0.68 \pm 0.02$ and ${\rm UCF}(\beta=4) =
0.66 \pm 0.01$.

\begin{figure}
\includegraphics[width=9cm,totalheight=6.5cm,angle=0]{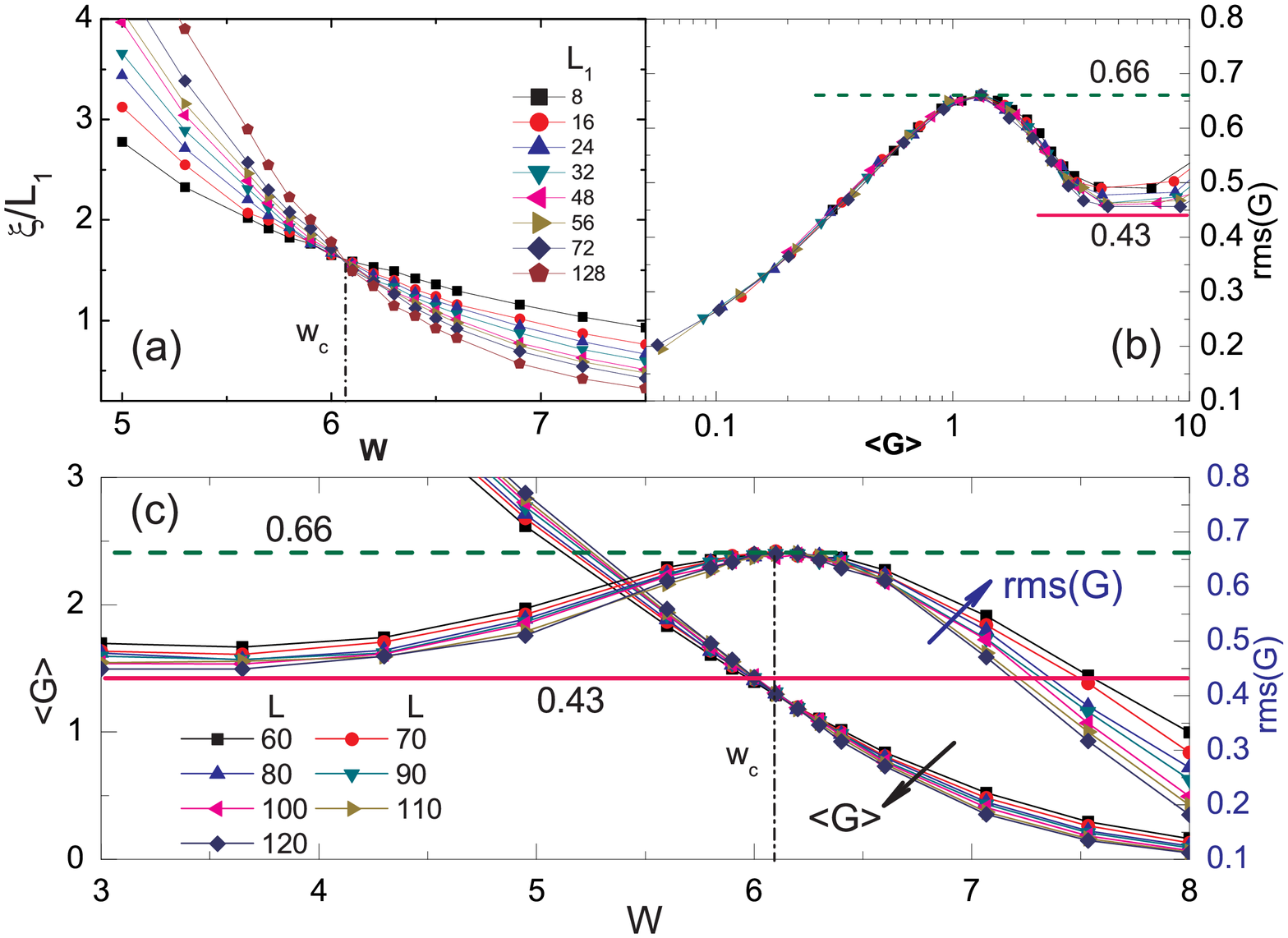}
\caption{(Color online) Localization length, average conductance,
conductance fluctuation of 2D symplectic systems for different $L$
with $E=2.3$ and $t_{so}=0.4$. (a). $\xi/L_1$ vs $W$. (b). $rms(G)$
vs $\left\langle G \right\rangle$ for different $L=60,70,...,120$.
(c). $rms(G)$ and $\left\langle G \right\rangle$ vs $W$.}
\label{fig5}
\end{figure}

Now we provide further evidence of the universal behavior of new
UCF. In 2D systems there can be a metal-insulator transition (MIT)
for $\beta=2,4$ but not for $\beta=1$. This somehow coincides with
our findings that there is a new UCF regime for $\beta=2,4$ but not
for $\beta=1$. To explore this correspondence further, we have
calculated the localization length of the 2D system using the
transfer matrix approach.\cite{mac} Here we calculate the
localization length $\xi$ for quasi-1D systems with fixed length
$1000000$ and different widths $L_1$ (see Fig.5a). Fig.5a depicts
the localization length $\xi/L_1$ vs disorder strength for
$\beta=4$. The intersection of different curves gives an estimate of
the critical disorder strength of MIT of the 2D system. Fig.5a shows
that for an infinite 2D system, there is a MIT around $W_c=6.1$. For
a mesoscopic 2D system, the critical region becomes a crossover
region around the same $W_c$ and it is in this region where the new
UCF is found. To see how our new UCF evolves with increasing of
system size $L$ we have calculated $rms(G)$ vs $W$ for finite 2D
systems with different sizes $L=50+10n$ where $n=1,2,...,7$. As
shown in Fig.5c, for a fixed $W$ that is beyond the crossover
regime, e.g., $W=7$, the fluctuation decreases as $L$ increases so
that $rms(G) \rightarrow 0$ at $L \rightarrow \infty$. Importantly,
the second plateau value (the new UCF) does not change with $L$. In
addition, both $\left\langle G \right\rangle$ and $rms(G)$ converge
at $W_c=6.1$ for different $L$. This means that when $L$ goes to
infinity we should have $rms(G) = c_\beta$ in the vicinity of
critical region where $c_\beta$ is the new UCF. This again suggests
that the new UCF is driven by MIT and is an universal quantity.
Finally $rms(G)$ vs $\left\langle G \right\rangle$ is plotted in
Fig.5b for different $L$ which shows the universal behavior that is
also independent of $L$. Similar behaviors were also observed for 2D
systems with $\beta=2$.

We have further calculated the third and fourth moments ($\gamma_1$
and $\gamma_2$) of conductance vs the average
conductance.\cite{ren2} For QD and 2D systems similar universal
features found in $rms(G)$ were also found for $\gamma_1$ and
$\gamma_2$ with different $\beta$.\cite{attach1} It is interesting
that in the localized regime with $\left\langle G \right\rangle<0.3$
the conductance distribution seems to be superuniversal that is
independent of dimensionality and symmetry.\cite{attach2}

In summary, we have carried out extensive simulations on conductance
fluctuations of quasi-1D, QD, and 2D mesoscopic systems for
orthogonal, unitary, and symplectic ensembles. Our results show that
in addition to the usual UCF in the diffusive regime there exists a
new UCF in the crossover regime between metallic and insulating
regimes for unitary and symplectic ensembles but not for the
orthogonal ensemble. We found that the conductance fluctuation
$rms(G)$ and higher order moments vs $\left\langle G \right\rangle$
are universal functions from diffusive to localized regimes which
depend only on symmetry index $\beta$ for quasi-1D, QD, and 2D
mesoscopic systems. In quasi-1D systems this universal function
agrees with the result from DMPK equation. Our analysis suggests
that this new UCF is driven by MIT in 2D systems.

\bigskip

{\bf Acknowledgments} This work was financially supported by a RGC
grant (HKU 704607P) from the government of HKSAR and LuXin Energy
Group. We wish to thank Prof. J.R. Shi for providing the code of
solving DMPK equation. Computer Centre of The University of Hong
Kong is gratefully acknowledged for the High-Performance Computing
assistance.

$^*$ Electronic address: jianwang@hkusua.hku.hk


\begin{thebibliography}{99}
\bibitem{foot4}
For 2d systems (in Fig.2d), we have shown data obtained from at
least five different set of parameters for each $\beta$.
\end{thebibliography}
\end{document}